\documentclass[conference]{IEEEtran}

\usepackage{color}
\usepackage{mathtools}
\usepackage{algorithmic}

\algsetup{linenosize=\small}
\usepackage[table,xcdraw]{xcolor}
\usepackage{multirow}
\usepackage[super]{nth}
\usepackage{graphicx}
\usepackage{caption}
\usepackage{subcaption}
\usepackage{textcomp}
\usepackage{xspace}
\usepackage{siunitx}
\usepackage{epsfig}
\usepackage{epstopdf}
\usepackage{soul}
\usepackage{url}
\usepackage{tablefootnote}
\usepackage{graphicx}
\usepackage[labelformat=simple]{subcaption}
\usepackage{epsfig} 
\usepackage{amsmath,amssymb}
\usepackage[linesnumbered,ruled]{algorithm2e}
\usepackage{makecell}
\usepackage{multirow}
\usepackage{adjustbox}
\usepackage[space]{cite}
\usepackage[labelformat=simple]{subcaption}

\usepackage[font=scriptsize]{subcaption}

\begin{document}
	
	%
	\title{Implications of Decentralized Q-learning Resource Allocation in Wireless Networks}
	
	\author{
		\IEEEauthorblockN{Francesc Wilhelmi, Boris Bellalta}
		\IEEEauthorblockA{
			Wireless Networking (WN-UPF)\\
			Universitat Pompeu Fabra\\
			Barcelona, Spain}
		\and
		\IEEEauthorblockN{Cristina Cano}
		\IEEEauthorblockA{WINE Group\\ Universitat Oberta de Catalunya\\
			Castelldefels, Spain}
		
		\and
		\IEEEauthorblockN{Anders Jonsson}
		\IEEEauthorblockA{
			Art. Int. and Mach. Learn. (AIML-UPF)\\
			Universitat Pompeu Fabra\\
			Barcelona, Spain}
	}
	
	\maketitle
	
	\begin{abstract}
		Reinforcement Learning is gaining attention by the wireless networking community due to its potential to learn good-performing configurations only from the observed results. In this work we propose a stateless variation of Q-learning, which we apply to exploit spatial reuse in a wireless network. In particular, we allow networks to modify both their transmission power and the channel used solely based on the experienced throughput. We concentrate in a completely decentralized scenario in which no information about neighbouring nodes is available to the learners. Our results show that although the algorithm is able to find the best-performing actions to enhance aggregate throughput, there is high variability in the throughput experienced by the individual networks.
		We identify the cause of this variability as the adversarial setting of our setup, in which the most played actions provide intermittent good/poor performance depending on the neighbouring decisions. We also evaluate the effect of the intrinsic learning parameters of the algorithm on this variability.
		
	\end{abstract}
	
	\IEEEpeerreviewmaketitle

	\section{Introduction}
	Reinforcement Learning (RL) has recently spread use in the wireless communications field to solve many kinds of problems such as Access Point (AP) association\cite{chen2010ap}, channel selection \cite{maghsudi2015channel} or transmit power adjustment \cite{maghsudi2015joint}, as it allows learning good-performing configurations only from the observed results. Among these, Q-learning has been applied to dynamic channel assignment in mobile networks in \cite{nie1999qlearning} and to automatic channel selection in Femto Cell networks in \cite{bennis2010q}. However, to the best of our knowledge, the case of a fully decentralized scenario where nodes do not have knowledge from each other, has not yet been considered.  
	
	In this work we propose a stateless variation of Q-learning in which nodes select the transmission power and channel to use solely based on their resulting throughput. We concentrate on a fully decentralized scenario where no information about the actions and resulting performance of the other nodes is available to the learners. Note that inferring the throughput of neighbouring nodes allocated to different channels is costly as periodic sensing in the other channels would then be needed. We aim to characterize the performance of Q-learning in such scenarios, obtaining insight on the most played actions (i.e., channel and transmit power selected) and the resulting performance. We observe that when no information about the neighbours is available to the learners, these will tend to apply selfish strategies that result in alternating good/poor performance depending on the actions of the others. In such scenarios, we show that the use of Q-learning allows each network to find the best-performing actions, though without reaching a steady solution. Note that achieving a steady solution in a decentralized environment relies in finding a Nash Equilibrium, a concept used in Game Theory to define a set of individual strategies that maximize the profits of each player in a non-cooperative game, regardless of the others' strategy. Formally, a set of best player actions $a^* = (a_1^*, ..., a_n^*) \in A$ leads to a Nash Equilibrium if $a_i^* \in B_i(a_{-i}^*), \forall i \in N$, where $B_i(a_{-i})$ is the best response to the others actions ($a_{-i}$). Thus, the consequences of not reaching a Nash Equilibrium can have an impact on performance variability.
	
	In addition, we look at the resulting performance in terms of throughput when varying several parameters intrinsic to the learning algorithm, which helps in understanding the interactions between the degree of exploration and learning rate, and the variability of the resulting performance. 
	
	The remaining of this document is structured as follows: Section \ref{section:system_model} introduces the simulation scenario and considerations. Then, Section \ref{section:qlearning} presents our Stateless variation of Q-learning and its practical implementation for the resource allocation problem in Wireless Networks (WNs). Simulation results are later discussed in Section \ref{section:performance_evaluation}. Finally, some final remarks are provided in Section \ref{section:conclusions}. 
	
	\section{System model}
	\label{section:system_model}	
	
	For the remainder of this work, we consider a scenario in which several WNs are placed in a 3D-map (with parameters described later in Section \ref{section:simulation_parameters}), each one formed by an Access Point (AP) transmitting to a single Station (STA) in downlink manner. 
	
	\subsection{Channel modelling}
	\label{section:channel_modelling}	
	
	Path-loss and shadowing effects are modelled using the log-distance model for indoor communications. The path-loss between WN $i$ and $j$ is given by	
	\begin{align}
	\text{PL}_{i,j} &= \text{P}_{{\rm tx},i} - \text{P}_{{\rm rx},j} = \nonumber \\ & = \text{PL}_0 + 10  \alpha_{\rm PL}  \log_{10}(d_{i,j}) + \text{G}_{{\rm s}} + \frac{d_{i,j}}{d_{{\rm obs}}} \text{G}_{{\rm o}}, \nonumber
	\end{align}
	where $\text{P}_{{\rm tx},i}$ is the transmitted power in dBm by WN $i$, $\text{P}_{{\rm rx},j}$ is the power in dBm received in WN $j$, $\text{PL}_0$ is the path-loss at one meter in dB, $\alpha_{\rm PL}$ is the path-loss exponent, $d_{i,j}$ is the distance between the transmitter and the receiver in meters, $\text{G}_{{\rm s}}$ is the shadowing loss in dB, and $\text{G}_{{\rm o}}$ is the obstacles loss in dB. Note that we include the factor $d_{{\rm obs}}$, which is the distance between two obstacles in meters. 
	
	\subsection{Throughput calculation}
	\label{section:throughput_calculation}
	
	By using the power received and the interference, we calculate the maximum theoretical throughput of each WN $i$ at time $t \in \{1,2 ...\}$ by using the Shannon Capacity.
	\begin{equation}
	\Gamma_{i,t} = B  \log_{2}(1 + \text{SINR}_{i, t}),
	\nonumber
	\label{eq:shannon_capacity}
	\end{equation}
	where $B$ is the channel bandwidth and the experienced Signal to Interference plus Noise Ratio (SINR) is given by:
	\begin{equation}
	\text{SINR}_{i,t} = \frac{\text{P}_{i,t}}{\text{I}_{i,t}+\text{N}},
	\nonumber
	\label{eq:sinr}
	\end{equation}
	where $\text{P}_{i,t}$ and $\text{I}_{i,t}$ are the received power and the sum of the interference at WN $i$ at time $t$, respectively, and N is the floor noise power. For each STA in a WN, the interference is considered to be the total power received from all the APs of the other coexisting WNs as if they were continuously transmitting. Adjacent channel interference is also considered in $\text{I}_{i,t}, i \in \{1,..,W\}$, where $W$ is the number of neighbouring WNs. We consider that the transmitted power leaked to adjacent channels is $20$ dBm lower for each channel separation.	
	
	\section{Decentralized Stateless Q-learning for enhancing Spatial Reuse in WNs}
	\label{section:qlearning}	
	Q-learning \cite{sutton1998reinforcement, watkins1992q} is an RL technique that enables an agent to learn the optimal policy to follow in a given environment. A set of possible states describing the environment and actions are defined in this model. In particular, an agent maintains an estimate of the expected long-term discounted reward for each state-action pair, and selects actions with the aim of maximizing it. The expected cumulative reward $\text{V}^\pi(s)$ is given by:
	\begin{equation}
	\label{eq:ql_reward_policy}
	\text{V}^\pi(s) = \lim_{N \rightarrow \infty} \mathop{\mathbb{E}}\Big(\sum_{t=1}^{N} r_t^\pi(s)\Big),
	\nonumber
	\end{equation}
	where $r_t^\pi(s)$ is the reward obtained at iteration $t$ after starting from state $s$ and by following policy $\pi$. Since the reward may easily get unbounded, a discount factor parameter ($\gamma < 1$) is used. The optimal policy $\pi^*$ that maximizes the total expected reward is given by the Bellman's Optimality Equation \cite{sutton1998reinforcement}:	
	\begin{equation}
	Q^*(s,a) = \mathbb{E} \Big\{r_{t+1} + \gamma \text{max}_{a'} Q^*(s_{t+1},a') | s_t = s, a_t = a\Big\}. \nonumber
	\end{equation}	
	Henceforth, Q-learning receives information about the current state-action tuple $(s_t,a_t)$, the generated reward $r_t$ and the next state $s_{t+1}$, in order to update the Q-table: 	
	\begin{equation}
	\hat{Q}(s_t,a_t)\leftarrow (1-\alpha_t) \hat{Q}(s_t,a_t) + \alpha_t \Big(r_t + \gamma \big(\underset{a'}{\text{max}}\hat{Q}(s_{t+1},a')\big)\Big),
	\nonumber
	\end{equation}	
	where $\alpha_t$ is the learning rate at time $t$, and $\underset{a'}{\text{max}}\hat{Q}(s_{t+1},a')$ is the best estimated value for the next state $s_{t+1}$. The optimal solution is 
	theoretically
	achieved with probability 1 if $\sum_{t=0}^{\infty} \alpha_t = \infty$, and $\sum_{t=0}^{\infty} \alpha_t^2 < \infty$, which satisfies that $\underset{t \rightarrow \infty}{\lim} \hat{Q}(s,a) = Q^*(s,a)$.
	Since we focus on a completely decentralized scenario where no information about the other nodes is available, the system can then be fully described by the set of actions and rewards.\footnote{We note that local information such as the observed instantaneous channel quality could be incorporated in the state definition. However, such a description of the system entails increased complexity.} Thus, we propose using a stateless variation of the original Q-learning algorithm. To implement decentralized learning to the resource allocation problem, we consider each WN to be an agent running Stateless Q-learning through an $\varepsilon$-greedy action-selection strategy, so that actions $a \in \mathcal{A}$ correspond to all the possible configurations that can be chosen with respect to the channel and transmit power. During the learning process we assume that WNs select actions sequentially, so that at each learning iteration, every agent takes an action in an ordered way. The order at which WNs choose an action at each iteration is randomly selected at the beginning of it. The reward after choosing an action is set as:
	\begin{equation}
	r_{i,t} = \frac{\Gamma_{i,t}}{\Gamma_i^*},
	\label{eq:reward_generation}
	\nonumber
	\end{equation}
	where $\Gamma_{i,t}$ is the experienced throughput at time $t$ by WN $i \in \{1,...,n\}$, being $n$ the number of WNs in the scenario, and $\Gamma_{i}^* = B \log_{2}(1 + \text{SNR}_{i})$ is WN $i$ maximum achievable throughput (i.e., when it uses the maximum transmission power and there is no interference).	Each WN applies the Stateless Q-learning as follows: 
	\begin{itemize}
		\item Initially, it sets the estimates of its actions $k \in \{1,...,K\}$ to 0: $\hat{Q}(a_k) = 0$.
		\item At each iteration, it applies an action by following the $\varepsilon$-greedy strategy, i.e., it selects the best-rewarding action with probability $1 - \varepsilon_t$, and a random one (uniformly distributed) the rest of the times.
		\item After choosing action $a_k$, it observes the generated reward (the relative experienced throughput), and updates the estimated value $\hat{Q}(a_k)$.
		\item Finally, $\varepsilon_t$ is updated to follow a decreasing sequence: $\varepsilon_t = \frac{\varepsilon_0}{\sqrt[]{t}}$.
	\end{itemize}	
	Note, as well, that the optimal policy cannot be derived for the presented scenario, but it can be approximated to enhance spatial reuse. This is due to the nature of the presented environment, as well as WNs decisions affect the others performance.
	Formally, the implementation details of Stateless Q-learning are described in Algorithm \ref{alg:qlearning}.
	The presented learning approach is intended to operate at the PHY level, allowing the operation of the current MAC-layer communication standards (e.g., in IEEE 802.11 WLANs, the channel access is governed by the CSMA/CA operation, so that Stateless Q-learning may contribute to improve spatial reuse at the PHY level).
	\begin{algorithm}
		\SetKwInOut{Input}{Input}
		\SetKwInOut{Output}{Output}		
		Function Stateless Q-learning $(\text{SINR},\mathcal{A})$\;
		\Input{SINR: Signal-to-Interference-plus-Noise Ratio sensed at the STA\\$\mathcal{A}$: set of possible actions in \{1, ..., K\}}
		\Output{$\overline{\Gamma}$: Mean throughput experienced in the WN}
		initialize: $t=0$, $\hat{Q}(a_k) = 0, \forall a_k \in \mathcal{A}$\\
		\While{active}
		{
			Select $a_k$  $\begin{cases}
			\underset{k=1,...,K}{\text{argmax }} \hat{Q}(a_k), & \text{with prob } 1 - \varepsilon\\
			i \sim \mathcal{U}(1, K), & \text{otherwise}
			\end{cases}$\\
			Observe reward $r_{a_k} = \frac{\Gamma_{a_k,t}}{\Gamma^*}$ \\
			$\hat{Q}(a_k) \leftarrow \hat{Q}(a_k) + \alpha \cdot \big(r_{a_k} + \gamma \cdot \max\hat{Q} - \hat{Q}(a_k)\big)$\\
			$\varepsilon_t \leftarrow \varepsilon_0 / \sqrt{t}$ \\	
			$ t \leftarrow t + 1$
		}
		\caption{Stateless Q-learning}
		\label{alg:qlearning}
	\end{algorithm}
	
	\section{Performance Evaluation}
	\label{section:performance_evaluation}	
	In this section we introduce the simulation parameters and describe the experiments.\footnote{The code used for simulations can be found at \url{https://github.com/wn-upf/Decentralized_Qlearning_Resource_Allocation_in_WNs.git} (Commit: eb4042a1830c8ea30b7eae3d72a51afe765a8d86).} Then, we show the main results.
	
	\subsection{Simulation Parameters}
	\label{section:simulation_parameters}
	According to \cite{bellalta2016ax}, a typical high-density scenario for residential buildings contains $0.0033 \text{APs}/\text{m}^3$. We then consider a map scenario with dimensions $10\times5\times10$ m containing 4 WNs that form a grid topology in which STAs are placed at the maximum possible distance from the other networks. This toy scenario allows us to study the performance of Stateless Q-learning in a controlled environment
	, which is useful to check the applicability of RL in WNs by only using local information \footnote{The analysis of the presented learning mechanisms in more congested scenarios is left as future work.}.
	We consider that the number of channels is equal to half the number of coexisting WNs, so that we can study a challenging situation regarding the spatial reuse. Table \ref{tbl:simulation_parameters} details the parameters used.	
	\begin{table}[h!]
		\centering
		\resizebox{\columnwidth}{!}{%
			\begin{tabular}{|l|l|}
				\hline
				\textbf{Parameter}             & \textbf{Value}                      \\ \hline
				Map size (m)                    & $10\times5\times10$                  \\ \hline
				Number of coexistent WNs                    & 4  \\ \hline
				APs/STAs per WN                    & 1 / 1                                     \\ \hline
				Distance AP-STA (m)     & $\sqrt{2}$                         \\ \hline
				Number of Channels              & 2                               \\ \hline
				Channel Bandwidth (MHz)         & 20                                    \\ \hline
				Initial channel selection model & Uniformly distributed 
				\\ \hline
				Transmit power values (dBm)                & \{5, 10, 15, 20\}                      \\ \hline				
				$\text{PL}_0$ (dB)                   & 5
				\\ \hline
				$\alpha_{\rm PL}$			& 4.4
				\\ \hline			
				$\text{G}_s$ (dB)                   & Normally distributed with mean 9.5 
				\\ \hline
				$\text{G}_o$ (dB)                  & Uniformly distributed with mean 30
				\\ \hline
				$d_{\rm obs}$  (meters between two obstacles)                 & 5                        
				\\ \hline
				Noise level (dBm)               & -100                                  \\ \hline
				Traffic model                   & Full buffer (downlink)             \\ \hline           
			\end{tabular}
		}
		\caption{Simulation parameters}
		\label{tbl:simulation_parameters}
	\end{table}
	
	\subsection{Optimal solution}
	\label{section:optimal_solution}	
	We first identify the optimal solutions that maximize: $i$) the aggregate throughput,  and $ii$) the proportional fairness, which is computed as the logarithmic sum of the throughput experienced by each WN, i.e., ${\rm PF} = \underset{k \in \mathcal{A}}{\rm max} \sum_i \log(\Gamma_{i,k})$. The optimal solutions are listed in Table \ref{tbl:optimal_configurations}. Note that, since the considered scenario is symmetric, there are two equivalent solutions. Note, as well, that in order to maximize the aggregate network throughput two of the WNs sacrifice themselves by choosing a lower transmit power. This result is then not likely to occur in an adversarial selfish setting.	
	\begin{table}[]
		\centering
		\resizebox{\columnwidth}{!}{%
			\begin{tabular}{|c|c|c|}
				\hline
				\textbf{WN id} & \textbf{\begin{tabular}[c]{@{}c@{}}Action that maximizes the \\ Aggregate Throughput\end{tabular}} & \textbf{\begin{tabular}[c]{@{}c@{}}Action that maximizes the\\ Proportional Fairness\end{tabular}} \\ \hline
				1 & 1 (2) & 7 (8) \\ \hline
				2 & 1 (2) & 8 (7) \\ \hline
				3 & 7 (8) & 7 (8) \\ \hline
				4 & 8 (7) & 8 (7) \\ \hline
			\end{tabular}	
		}		
		\caption{Optimal configurations (action indexes) to achieve the maximum network throughput and prop. fairness, resulting in 1124 Mbps and 891 Mbps, respectively. In parenthesis the analogous solution is shown. Actions indexes range from 1 to 8 are mapped to \{channel number, transmit power (dBm)\}: \{1,5\}, \{2,5\}, \{1,10\}, \{2,10\}, \{1,15\}, \{2,15\},\{1,20\} and \{2,20\}, respectively.}
		\label{tbl:optimal_configurations}
	\end{table}
	
	\subsection{Input Parameters Analysis}
	\label{section:practical_analysis}	
	We first analyse the effects of modifying $\alpha$ (the learning rate), $\gamma$ (the discount factor) and $\varepsilon_0$ (the initial exploration coefficient of the $\varepsilon$-greedy update rule) with respect to the achieved network throughput. We run simulations of $10000$ iterations and capture the results of the last $5000$ iterations to ensure that the initial transitory phase has ended. Each simulation is repeated $100$ times for averaging purposes. 
	
\begin{figure}[]
		\centering
		\epsfig{file=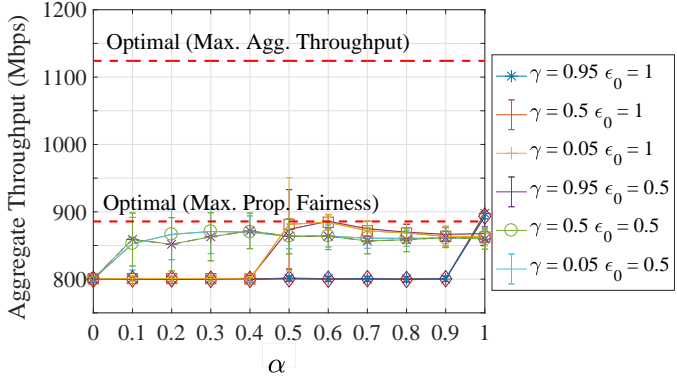, width=9.5cm}
		\caption{Effect of $\alpha, \gamma$ and $\epsilon_0$ in the average aggregate throughput ($100$ simulation runs per sample).}
		\label{fig:ql_alpha_gamma_epsilon_evaluation}
	\end{figure}

	Figure \ref{fig:ql_alpha_gamma_epsilon_evaluation} shows the average aggregate throughput achieved for each of the proposed combinations. It can be observed that the best results with respect to the aggregate throughput, regarding both average and variance, are achieved when $\alpha = 1$, $\gamma = 0.95$ and $\varepsilon_0 = 1$. This means that for achieving the best results (i.e., high average aggregate throughput and low variance), the immediate reward of a given action must be considered rather than any previous information ($\alpha = 1$). We see that the difference between the pay-off offered by the best action and the current one must also be high ($\gamma = 0.95$). In addition, exploration must be highly boosted at the beginning ($\varepsilon_0=1$). For this setting, the resulting throughput ($902.739$ Mbps) represents $80.29$\% of the one provided by the optimal configuration that maximizes the aggregate throughput (shown in Table \ref{tbl:optimal_configurations}). Regarding proportional fairness, the algorithm's resulting throughput is only $1.32$\% higher than the optimal. 
	
\begin{figure}[t!]
		\centering
		\begin{subfigure}[(a)]{0.32\textwidth}
			\includegraphics[width=\textwidth]{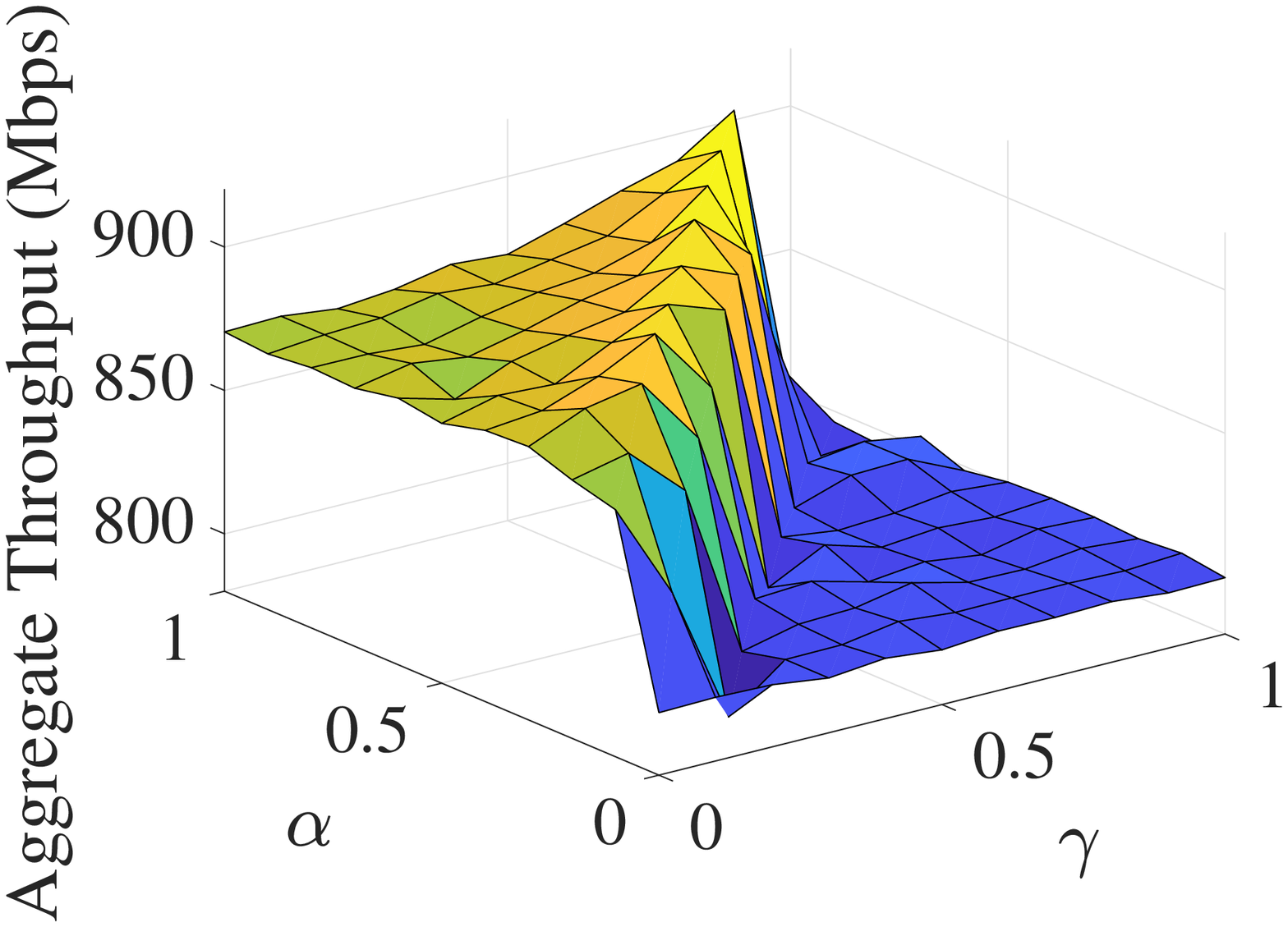}
			\caption{Av. aggregate throughput}
			\label{fig:alpha_vs_gamma_avg_tpt_sqrt_epsilon}
		\end{subfigure}
		\begin{subfigure}[(b)]{0.32\textwidth}
			\includegraphics[width=\textwidth]{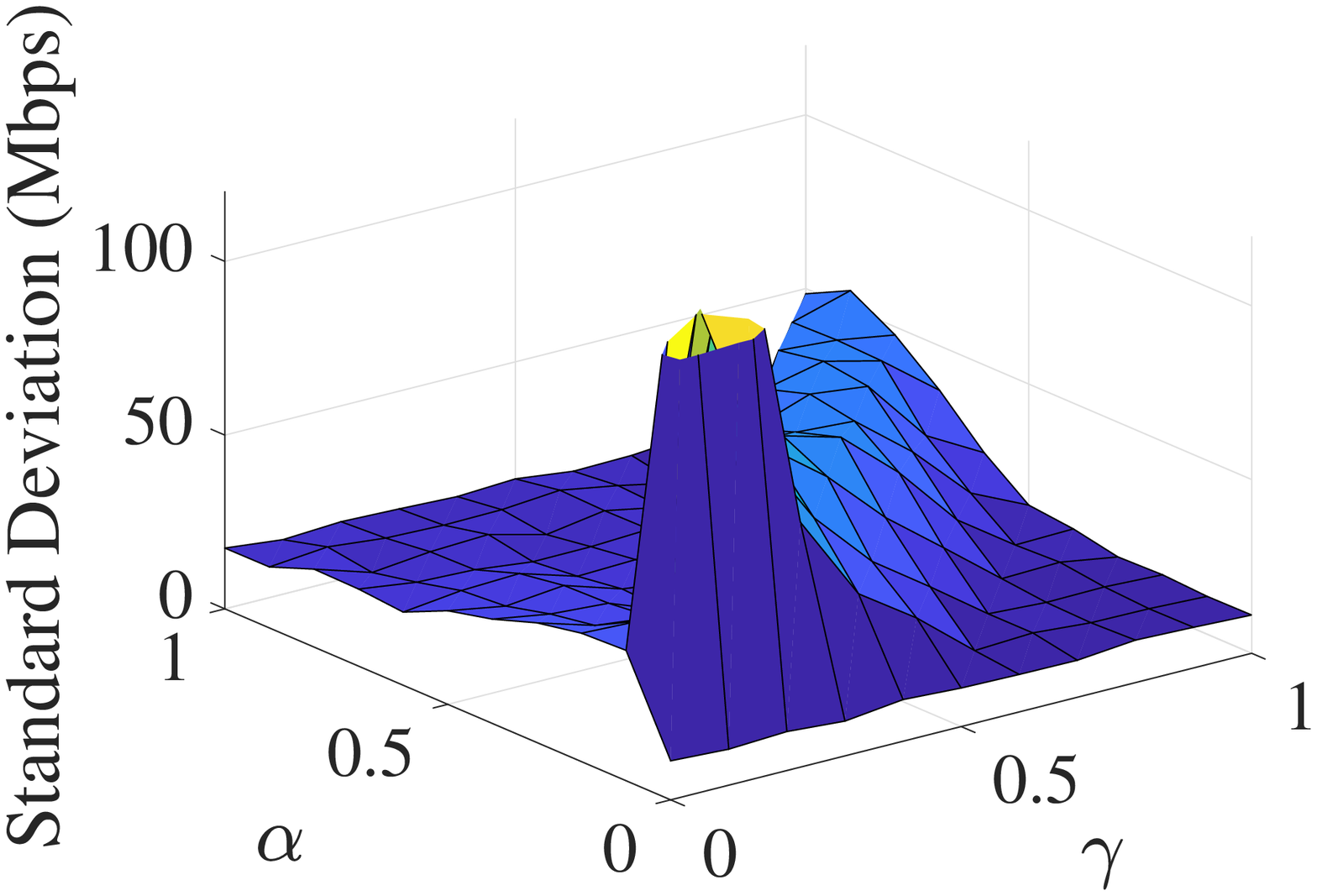}
			\caption{Standard deviation}
			\label{fig:alpha_vs_gamma_std_sqrt_epsilon}
		\end{subfigure}
		\caption{Evaluation of $\alpha$ and $\gamma$.}
		\label{fig:ql_alpha_vs_gamma}
	\end{figure}

	We also evaluate the relationship between different values of $\alpha$ and $\gamma$ in the average aggregate throughput and standard deviation (shown in Figure \ref{fig:ql_alpha_vs_gamma}). We observe a remarkably higher aggregate throughput when $\alpha > \gamma$. We also see that the variability between different simulation runs is much lower when the average throughput is higher. Additionally, we note a peak in the standard deviation when $\gamma \approx \alpha$ and $\gamma > \alpha$. 	
	
\begin{figure}[]
		\centering
		\begin{subfigure}[b]{0.225\textwidth}
			\includegraphics[width=\textwidth]{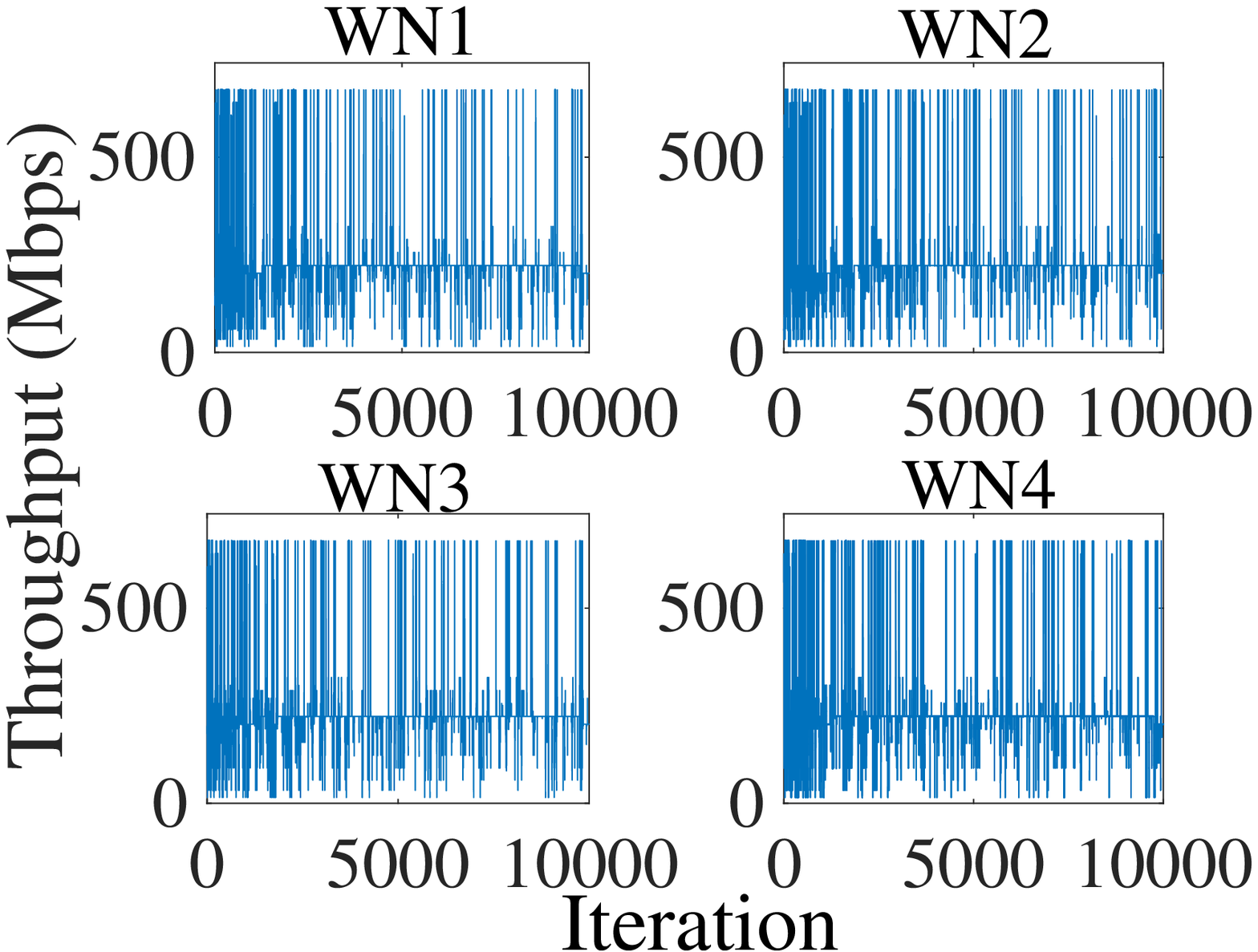}
			\caption{$\varepsilon_0=1, \alpha=1, \gamma=0.95$}
			\label{fig:e_1_a1_g095_ind_tpt}
		\end{subfigure}
		\begin{subfigure}[b]{0.225\textwidth}
			\includegraphics[width=\textwidth]{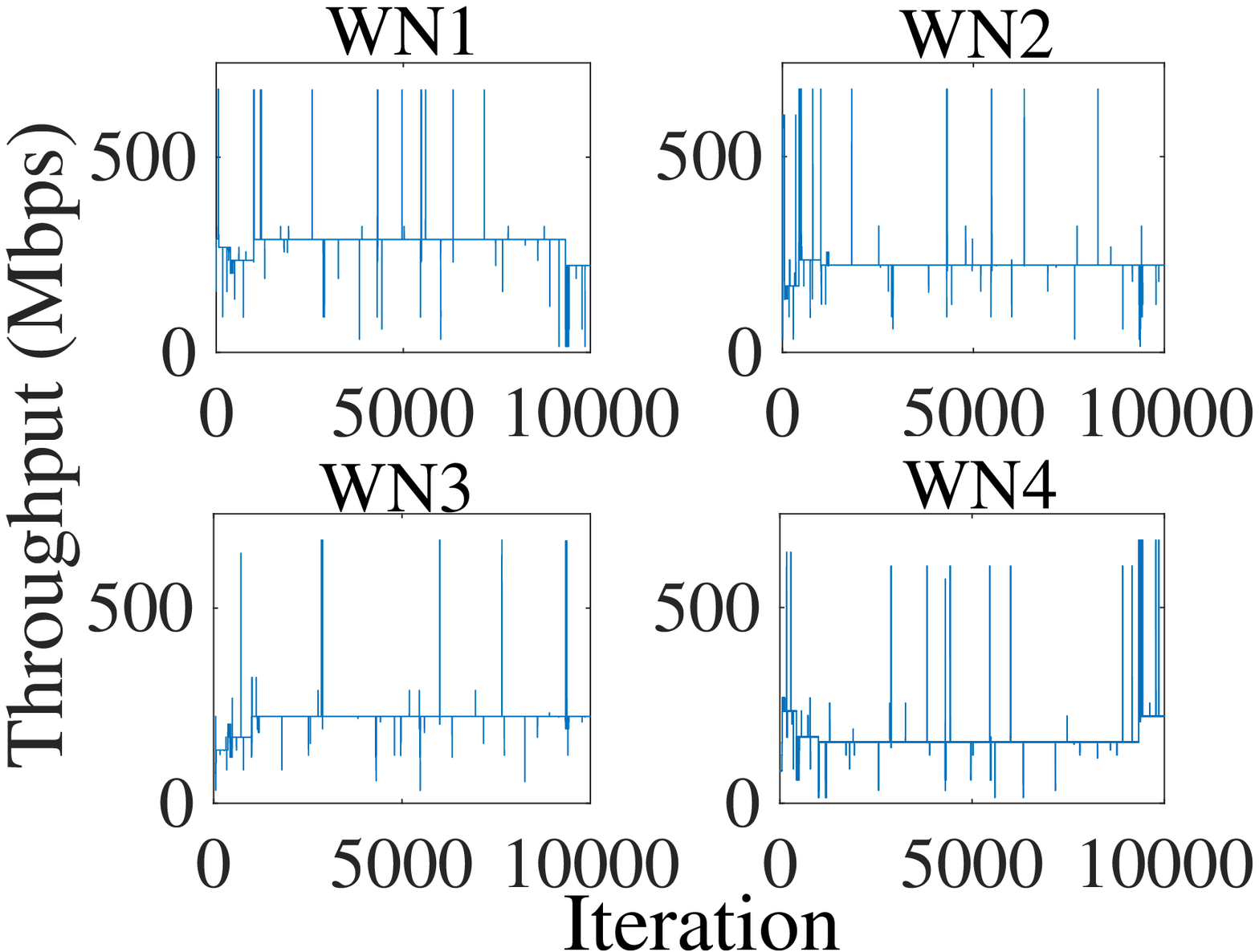}
			\caption{$\varepsilon_0=0.1, \alpha=1, \gamma=0.95$}
			\label{fig:e_1_a_1_g_095_ind_tpt}
		\end{subfigure}
		\begin{subfigure}[b]{0.225\textwidth}
			\includegraphics[width=\textwidth]{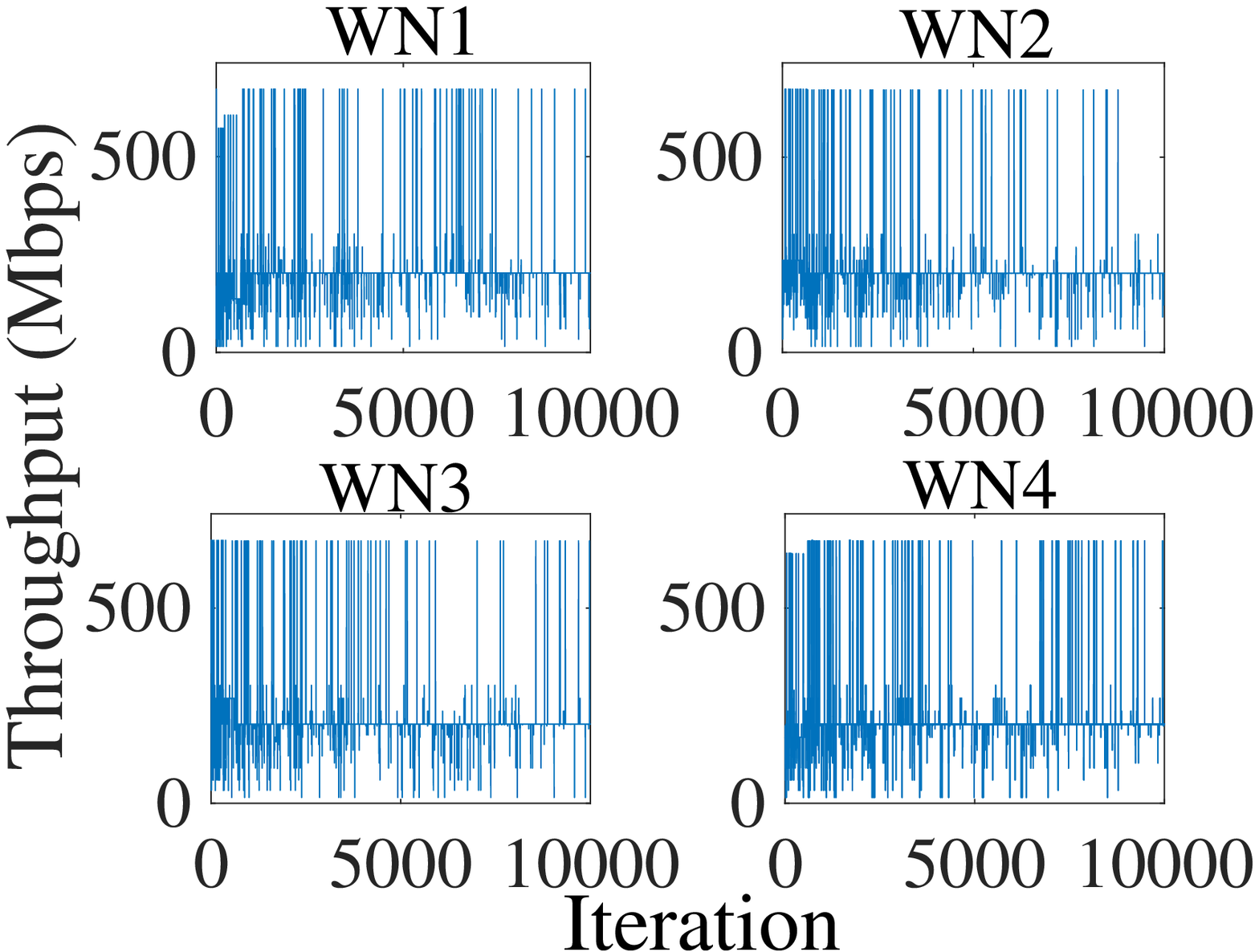}
			\caption{$\varepsilon_0=1, \alpha=0.1, \gamma=0.05$}
			\label{fig:e_1_a_01_g_005_ind_tpt}
		\end{subfigure}
		\begin{subfigure}[b]{0.225\textwidth}
			\includegraphics[width=\textwidth]{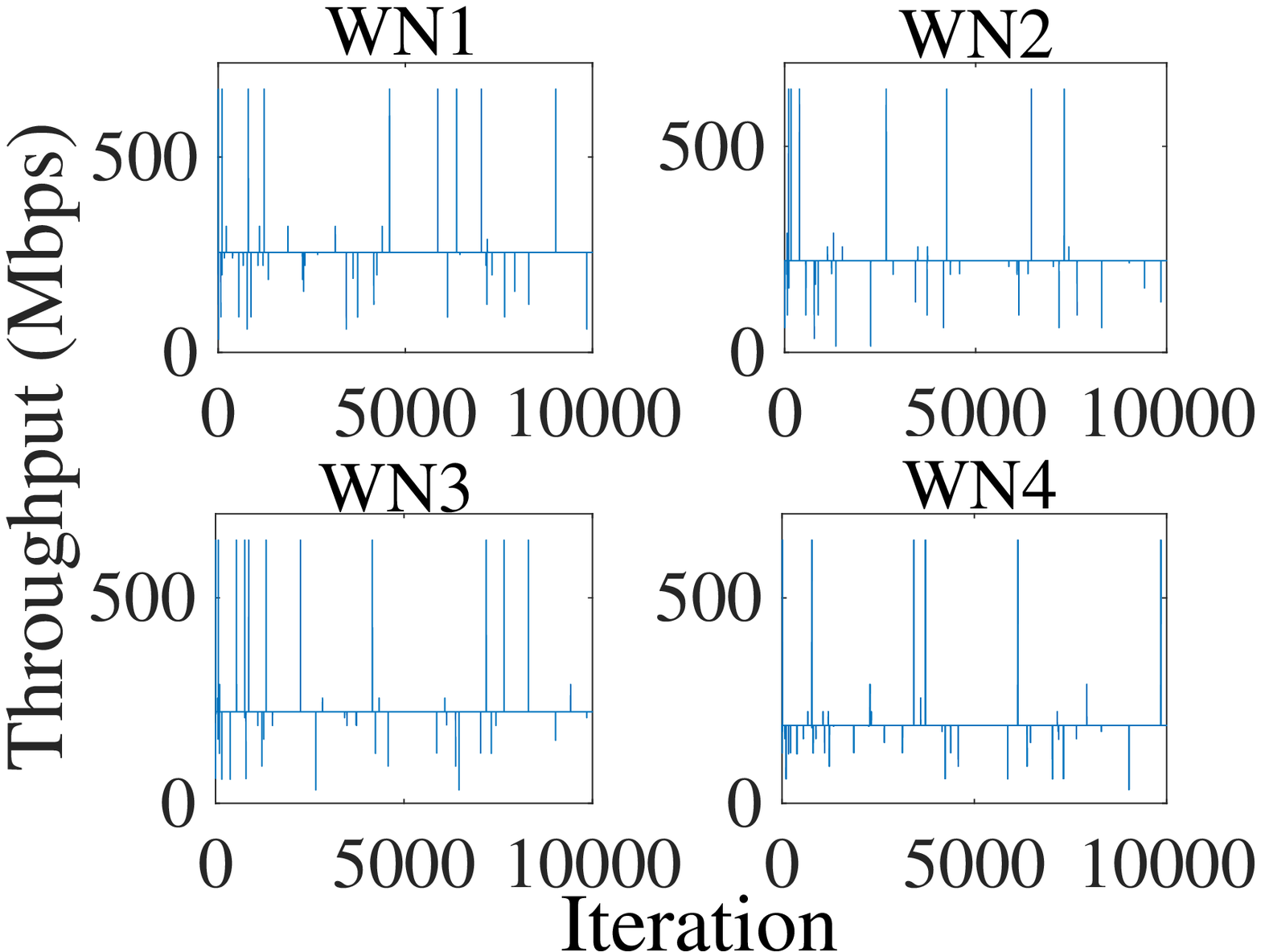}
			\caption{$\varepsilon_0=0.1, \alpha=0.1, \gamma=0.05$}
			\label{fig:e_01_a_01_g_005_ind_tpt}
		\end{subfigure}
		\caption{Individual throughput experienced by each WN during a single simulation run for different $\varepsilon_0$, $\alpha$ and $\gamma$.}
		\label{fig:ql_params_eval_individual_tpt}
	\end{figure}

\begin{figure}[]
		\centering
		\begin{subfigure}[b]{0.225\textwidth}
			\includegraphics[width=\textwidth]{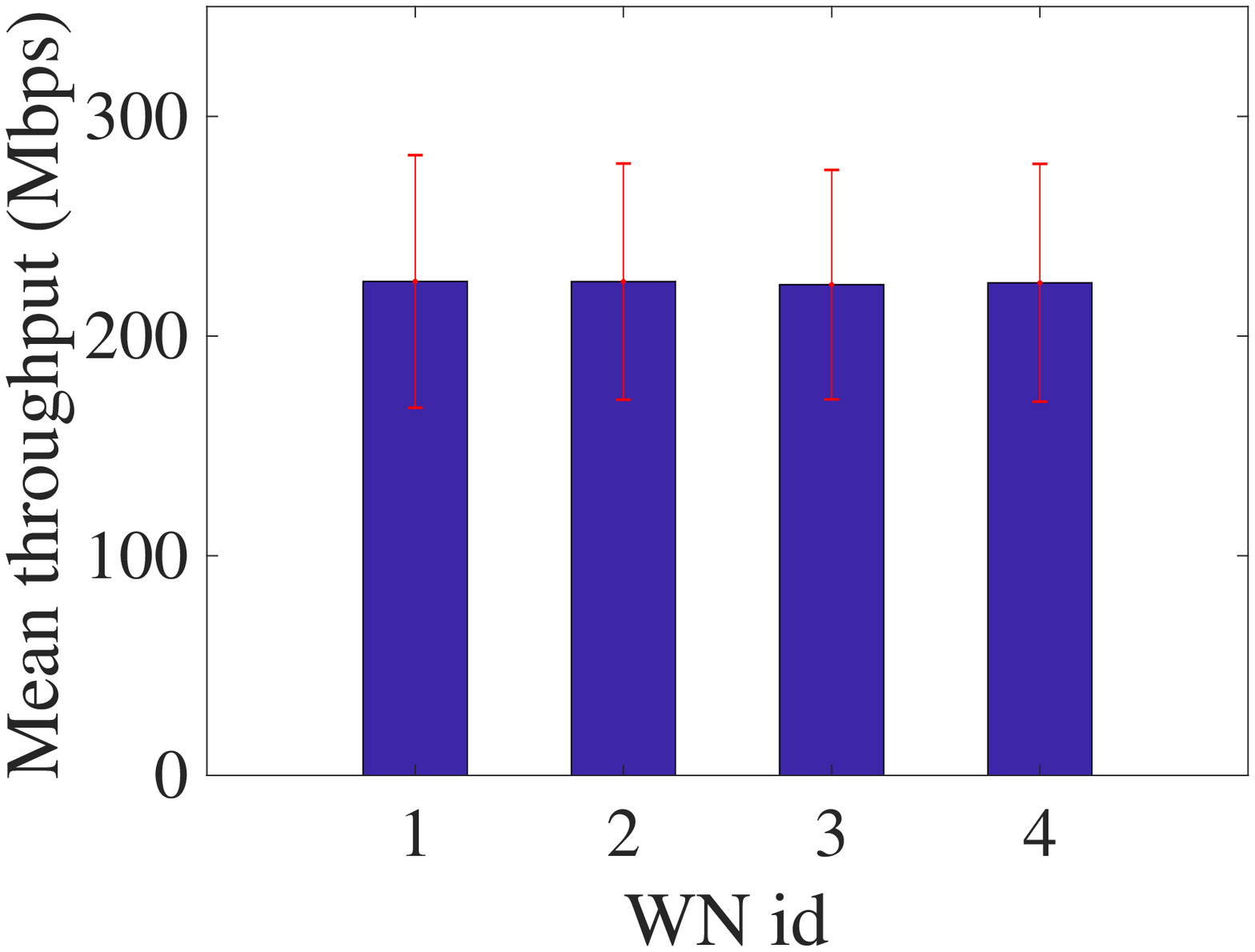}
			\caption{$\varepsilon_0=1, \alpha=1, \gamma=0.95$}
			\label{fig:e1_a_1_g_0.95_avg_tpt}
		\end{subfigure}
		\begin{subfigure}[b]{0.225\textwidth}
			\includegraphics[width=\textwidth]{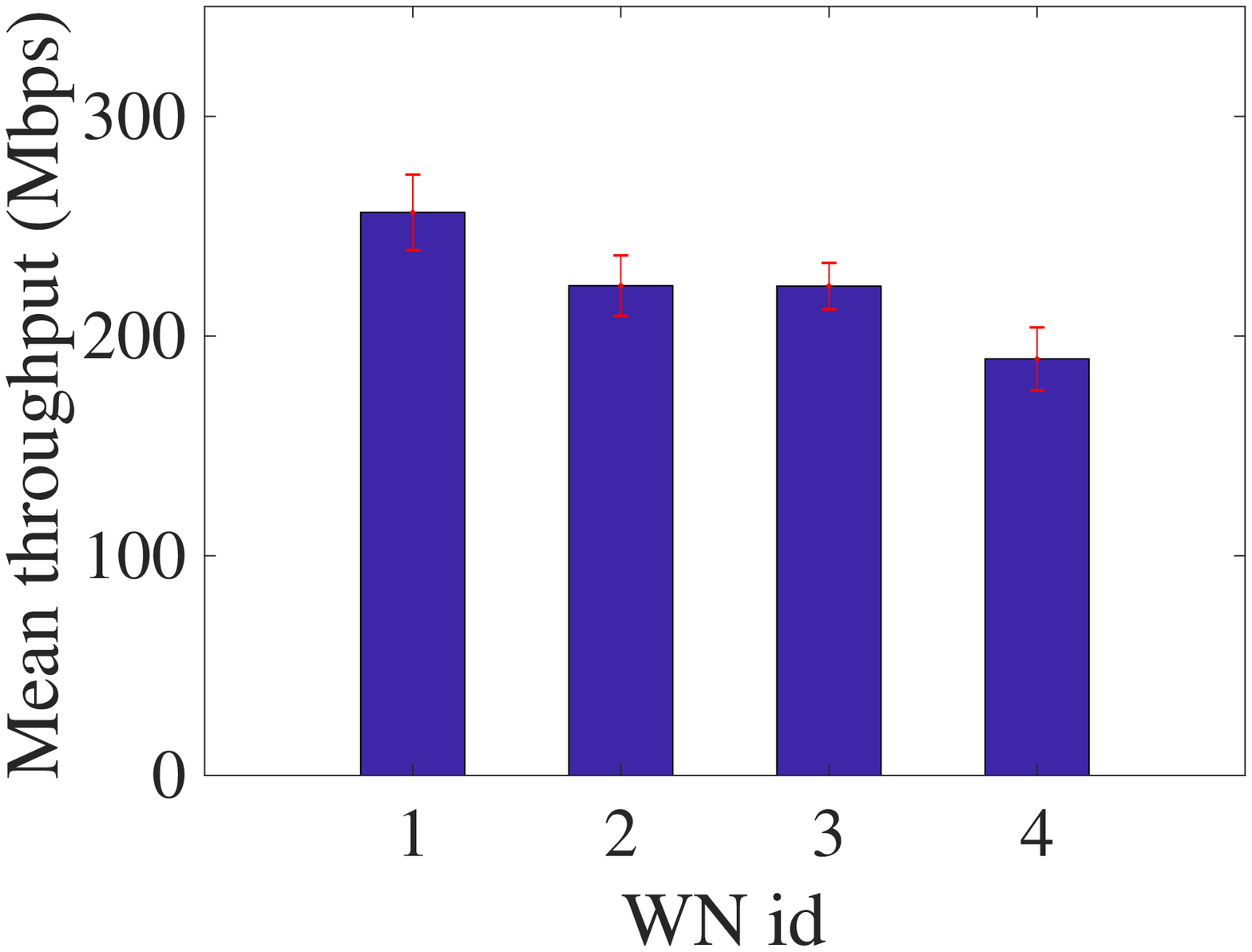}
			\caption{$\varepsilon_0=0.1, \alpha=1, \gamma=0.95$}
			\label{fig:e_1_a_1_g_0.95_avg_tpt}
		\end{subfigure}
		\begin{subfigure}[b]{0.225\textwidth}
			\includegraphics[width=\textwidth]{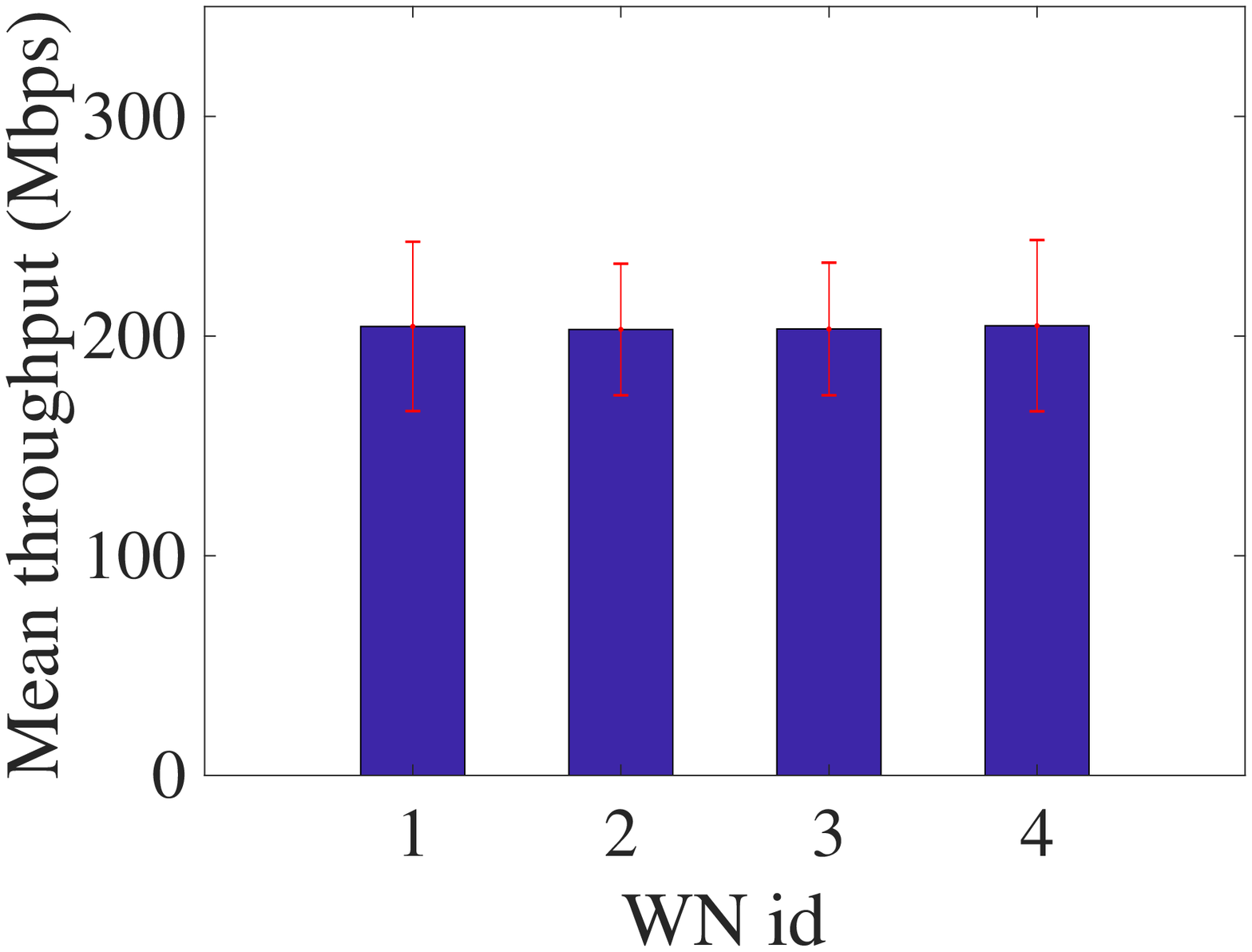}
			\caption{$\varepsilon_0=1, \alpha=0.1, \gamma=0.05$}
			\label{fig:e_1_a_0.1_g_0.05_avg_tpt}
		\end{subfigure}
		\begin{subfigure}[b]{0.225\textwidth}
			\includegraphics[width=\textwidth]{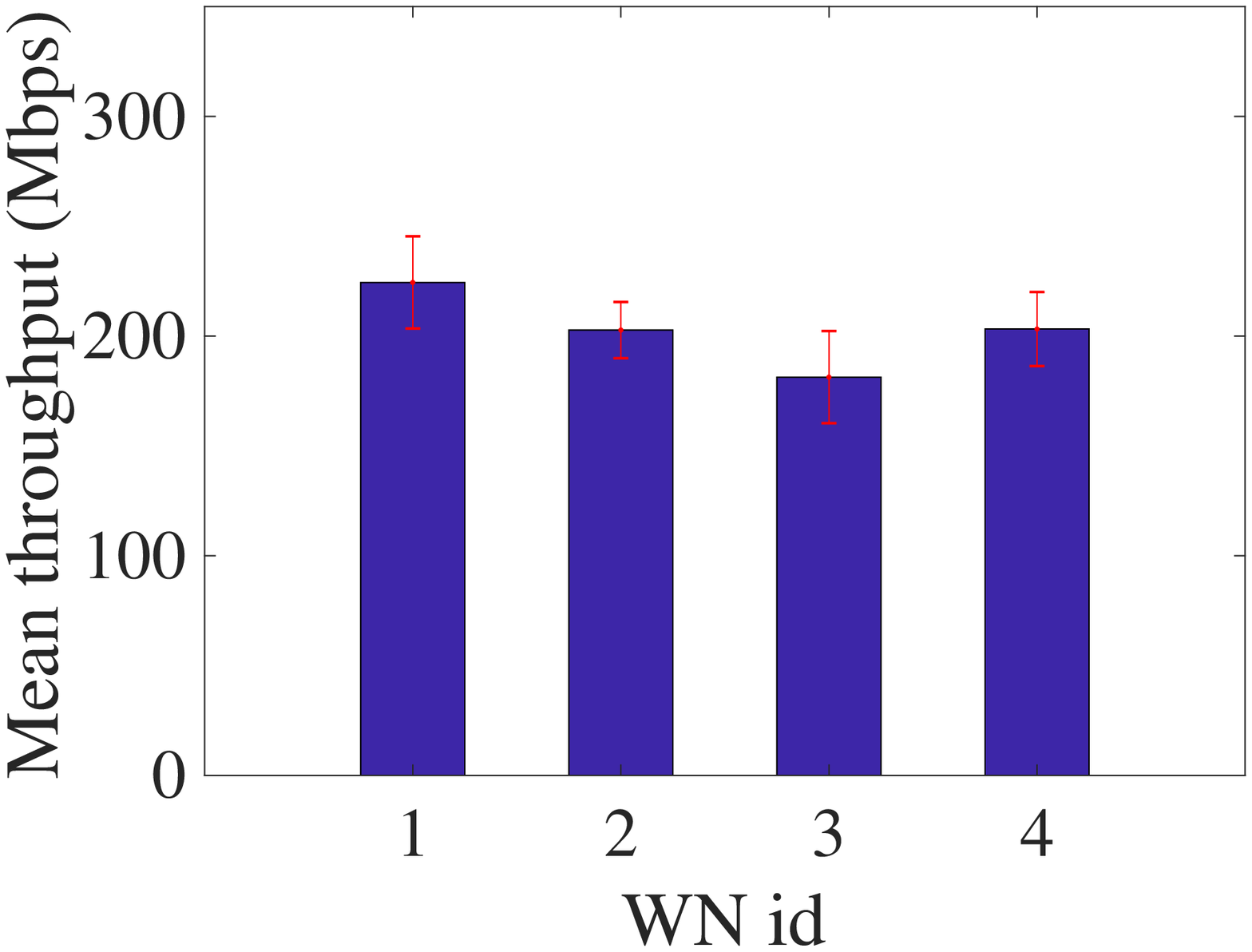}
			\caption{$\varepsilon_0=0.1, \alpha=0.1, \gamma=0.05$}
			\label{fig:e_0.1_a_0.1_g_0.05_avg_tpt}
		\end{subfigure}
		\caption{Average throughput experienced by each WN during the last 5000 iterations of a total of 10000 iterations (in a single simulation run) and for different $\varepsilon_0$, $\alpha$ and $\gamma$.}
		\label{fig:ql_params_eval_average_tpt}
	\end{figure}

\begin{figure}[]
		\centering
		\begin{subfigure}[b]{0.225\textwidth}
			\includegraphics[width=\textwidth]{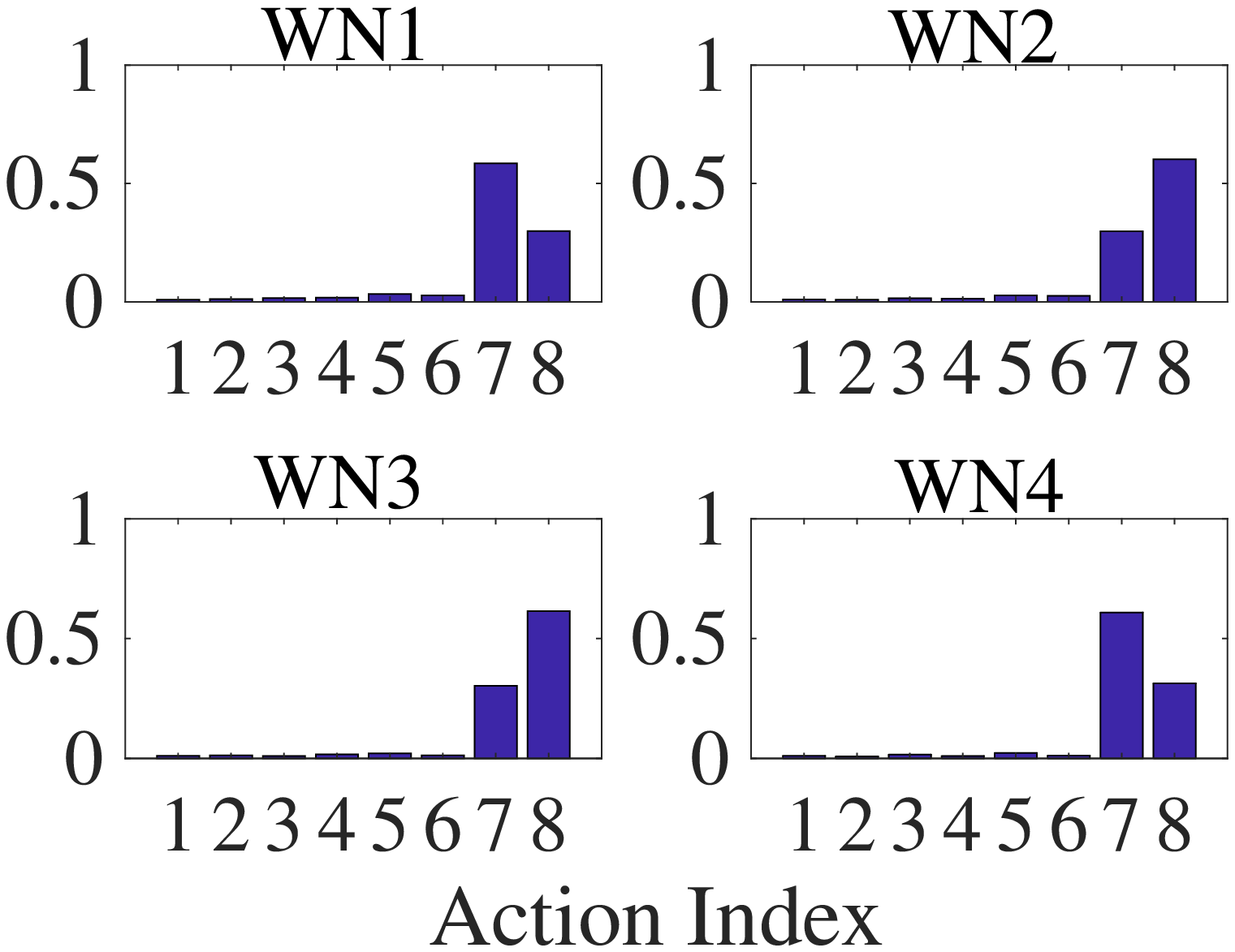}
			\caption{$\varepsilon_0=1, \alpha=1, \gamma=0.95$}
			\label{fig:e_1_a1_g095}
		\end{subfigure}
		\begin{subfigure}[b]{0.225\textwidth}
			\includegraphics[width=\textwidth]{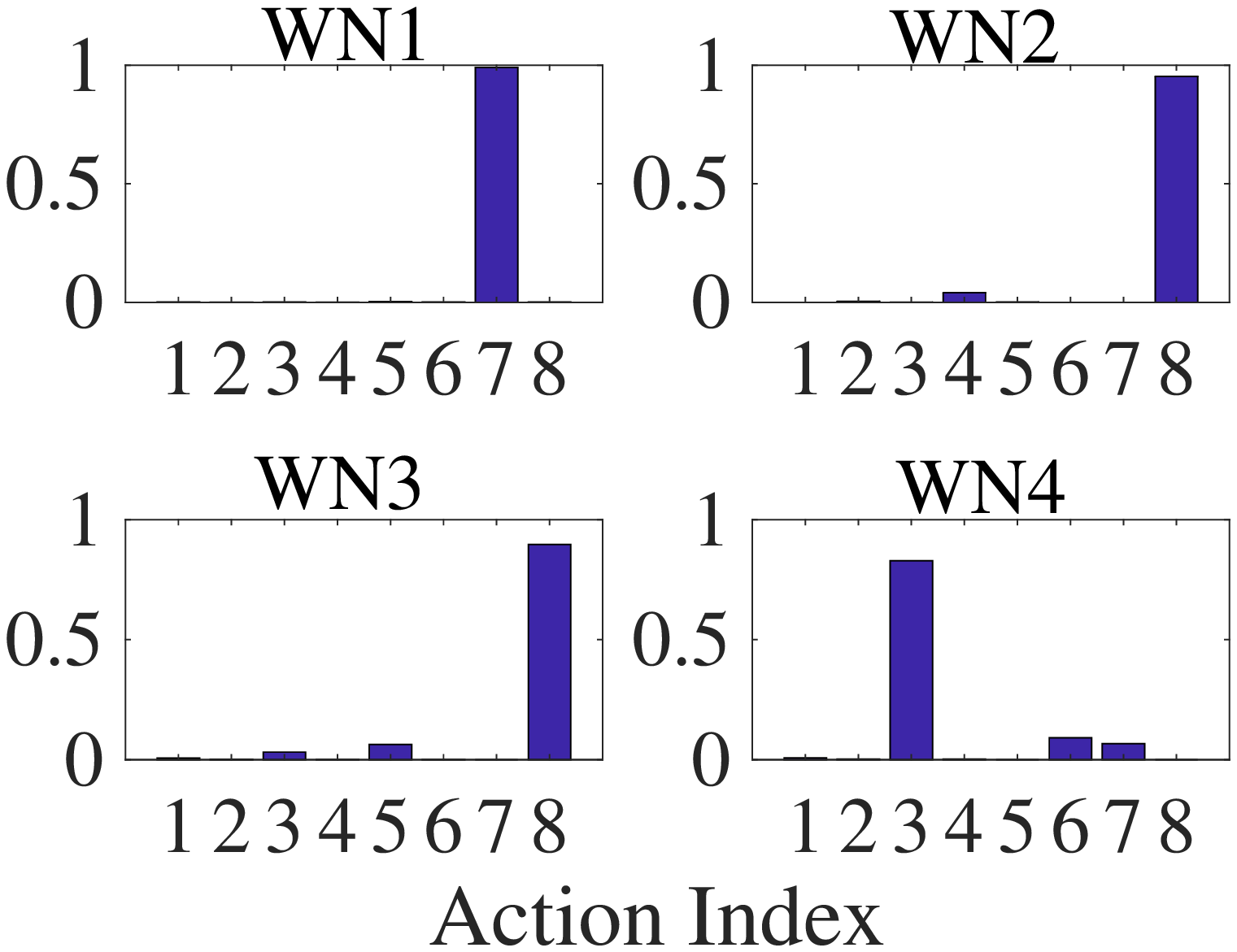}
			\caption{$\varepsilon_0=0.1, \alpha=1, \gamma=0.95$}
			\label{fig:e_1_a_1_g_095}
		\end{subfigure}
		\begin{subfigure}[b]{0.225\textwidth}
			\includegraphics[width=\textwidth]{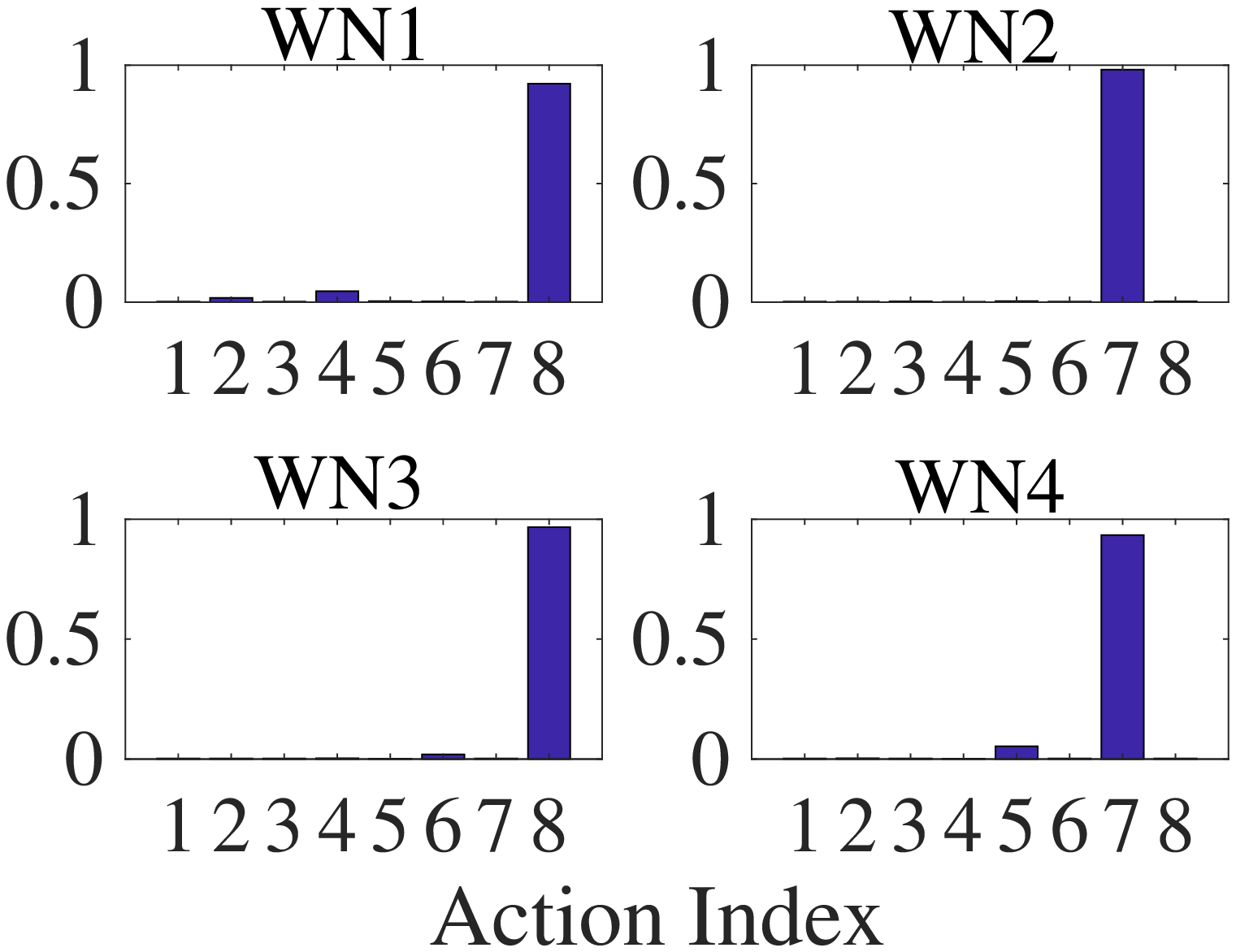}
			\caption{$\varepsilon_0=1, \alpha=0.1, \gamma=0.05$}
			\label{fig:e_1_a_01_g_005}
		\end{subfigure}
		\begin{subfigure}[b]{0.225\textwidth}
			\includegraphics[width=\textwidth]{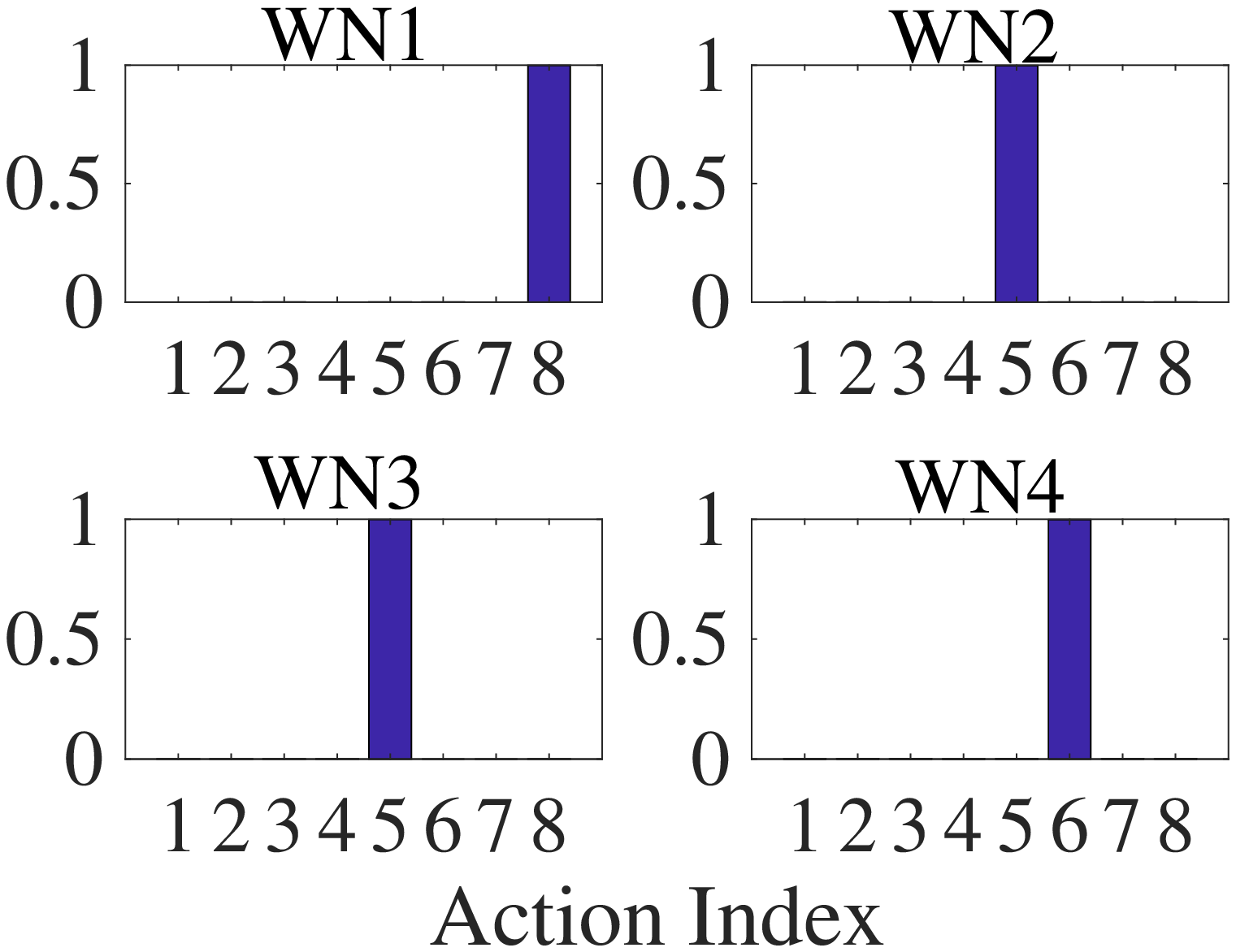}
			\caption{$\varepsilon_0=0.1, \alpha=0.1, \gamma=0.05$}
			\label{fig:e_01_a_01_g_005}
		\end{subfigure}
		\caption{Probability of choosing the different actions at each WN for a single (10000 iterations) simulation run and different $\varepsilon_0$, $\alpha$ and $\gamma$ values}
		\label{fig:ql_params_eval_actions_prob}
	\end{figure}

	To further understand the effects of modifying each of the aforementioned parameters, we show for different $\varepsilon_0$, $\alpha$ and $\gamma$: $i$) the individual throughput experienced by each WN during the total $10000$ iterations of a single simulation run (Figure \ref{fig:ql_params_eval_individual_tpt}), $ii$) the average throughput experienced by each WN for the last 5000 iterations, also for a single simulation run (Figure \ref{fig:ql_params_eval_average_tpt}), and $iii$) the probability of choosing each action at each WN (Figure \ref{fig:ql_params_eval_actions_prob}). We observe the following aspects:
	\begin{itemize}
		\item In Figure \ref{fig:ql_params_eval_individual_tpt} a high variability of the throughput experienced by each WN can be observed, specially if $\epsilon_0$ is high (as in Figures \ref{fig:e_1_a1_g095_ind_tpt}, \ref{fig:e_1_a_01_g_005_ind_tpt}). A high degree of exploration allows WNs to discover changes in the resulting performance of their actions due to the activity of the other nodes, which at the same time generates more variability (WN adapt to changes in the environment).
		\item Despite the variability generated, we obtain fairer results for high $\epsilon_0$ (Figure \ref{fig:ql_params_eval_average_tpt}). Henceforth, there is a relationship between the variability generated and the average throughput fairness.
		\item Finally, in Figures \ref{fig:e_1_a1_g095} and \ref{fig:e_1_a_01_g_005} we observe that for the former, there are two favourite actions that are being played the most, but for the latter there is only one preferred action. The lower the learning rate ($\alpha$), and consequently the discount factor ($\gamma$), the higher the probability of choosing a unique action, which results to be the one that provided the best performance in the past. The opposite occurs for higher $\alpha$ and $\gamma$ values, since giving more importance to the immediate reward allows for a reaction only to the recently-played actions of the neighbouring nodes: the algorithm is short-sighted. 
	\end{itemize}
	
	\section{Conclusions }
	\label{section:conclusions}		
	Decentralized Q-learning can be used to improve spatial reuse in dense wireless networks, enhancing performance as a result of exploiting the most rewarding actions. We have shown in this article, by means of a toy scenario, that Stateless Q-learning in particular allows finding good-performing configurations that achieve close-to-optimal (in terms of throughput maximization and proportional fairness) solutions. 
	
	However, the competitiveness of the presented fully-decentralized environment involves the non-existence of a Nash Equilibrium. Thus, we have also identified high variability in the experienced individual throughput due to the constant changes of the played actions, motivated by the fact that the reward generated by each action changes according to the opponents' ones. We have evaluated the impact of the parameters intrinsic to the learning algorithm on this variability showing that it can be reduced by decreasing the exploration degree and learning rate. The individual reduction on the throughput variability occurs at the expense of losing aggregate performance.
	
	This variability can potentially result in negative effects on the overall WN's performance. The effects of such a fluctuation in higher layers of the protocol stack can have severe consequences depending on the time scale at which they occur. For example, noticing high throughput fluctuations may trigger congestion recovery procedures in TCP (Transmission Control Protocol), which would harm the experienced performance. 
	
	We left for future work to further extend the decentralized approach in order to find collaborative algorithms that allow the neighbouring WNs to reach an equilibrium that grants acceptable individual performance. Acquiring any kind of knowledge about the neighbouring WNs is assumed to solve the variability issues arisen from decentralization. This information may be directly exchanged or inferred from observations. 
	Furthermore, other learning approaches are intended to be analysed in the future for performance comparison in the resource allocation problem. 
	
	\section*{Acknowledgment}
	This work has been partially supported by the Spanish Ministry of Economy and Competitiveness under the Maria de Maeztu Units of Excellence Programme (MDM-2015-0502), and by the European Regional Development Fund under grant TEC2015-71303-R (MINECO/FEDER). 
	

\end{document}